# Ultrathin MgB$_2$ films fabricated on Al$_2$O$_3$ substrate by hybrid physical-chemical vapor deposition with high $T_c$ and $J_c$

Yuhao Zhang[1], Zhiyuan Lin[1], Qian Dai, Dongyao Li, Yinbo Wang,

Yan Zhang, Yue Wang[*] and Qingrong Feng[†]

*State Key Laboratory for Mesoscopic Physics, Applied Superconductivity Center, and School of Physics, Peking University, Beijing 100871, People's Republic of China*

[1]These authors contributed equally to this work.

[*] yue.wang@pku.edu.cn
[†] qrfeng@pku.edu.cn

## Abstract

Ultrathin MgB$_2$ superconducting films with a thickness down to 7.5 nm are epitaxially grown on (0001) Al$_2$O$_3$ substrate by hybrid physical-chemical vapor deposition method. The films are phase-pure, oxidation-free and continuous. The 7.5 nm thin film shows a $T_c(0)$ of 34 K, which is so far the highest $T_c(0)$ reported in MgB$_2$ with the same thickness. The critical current density of ultrathin MgB$_2$ films below 10 nm is demonstrated for the first time as $J_c \sim 10^6$ A·cm$^{-2}$ for the above 7.5 nm sample at 16 K. Our results reveal the excellent superconducting properties of ultrathin MgB$_2$ films with thicknesses between 7.5 and 40 nm on Al$_2$O$_3$ substrate.

# I. Introduction

Over the last few years, the potential use of superconductors in electronic devices has attracted increasing attention as the superconducting devices show extremely low noise performance [1]. Among these devices, hot electron bolometers (HEBs) [2] and superconducting single-photon detectors (SSPDs) [3] are expected to be two promising applications with high performance. SSPDs based on NbN have been demonstrated experimentally with the detection efficiency as high as ~ 57% [4]. However, their working temperature and response time are limited by the critical temperature ($T_c$ ~ 16 K) and electron-phonon coupling time ($\tau_0$ ~ 0.06 ns) of the NbN. As a novel superconductor with $T_c$ ~ 40 K and $\tau_0$ ~ 0.002 ns [5], $MgB_2$ has considerable potential for use in these devices since it allows higher operation temperature and faster response time than NbN [6].

To fabricate HEBs and SSPDs, ultrathin superconducting $MgB_2$ films must be fabricated to guarantee the films are not affected by the thermal energy of the stimulated electron [7]. Thus, the growth of ultrathin $MgB_2$ films with a high transition temperature, high current-carrying capability and continuous surface is highly desirable. Due to difficulties in the synthesis, there are only a few reports on ultrathin $MgB_2$ films so far [8, 9, 10, 11]. Among these reports, it is worth noting that there is a large variance in the $T_c$ of the films grown on different substrates and by different experimental techniques. In the work of Shimakage *et al* [8] and Shibata *et al* [9], ultrathin $MgB_2$ films on $Al_2O_3$ substrate were fabricated by co-evaporation and molecular beam epitaxy respectively, showing that for 10 nm thick films the $T_c$ is only about 20 K, i.e., nearly half that of the $T_c$ of bulk $MgB_2$. Whereas, in the work of Wang *et al* [10] and Zhuang *et al* [11], where the hybrid physical-chemical vapor deposition (HPCVD) of ultrathin $MgB_2$ films on SiC substrate was performed, it was shown that the $T_c$ of the films with the same thickness of 10 nm can retain a high value of about 35 K. Whether the significantly lower $T_c$ of the films reported in [8] and [9] is due to the $Al_2O_3$ substrate or the growth techniques still needs to be clarified. This issue is important since, as a key parameter, the $T_c$ value of the ultrathin $MgB_2$ film will determine the prospects for the application of $MgB_2$ in HEBs or SSPDs and therefore we need to figure out which are decisive factors in determining the $T_c$ of ultrathin $MgB_2$ films. Moreover, it is known that the $Al_2O_3$ substrate is one of the most common and widely used substrates in the epitaxial growth of $MgB_2$ films and the manufacture of related superconducting devices. Hence, if it was experimentally found that one could also obtain a high $T_c$ for ultrathin $MgB_2$ films on $Al_2O_3$ substrate, the promise for $MgB_2$ in applications such as SSPDs could have a more solid basis and attract more interest.

To address the above issue, we have conducted the HPCVD synthesis of ultrathin $MgB_2$ films on $Al_2O_3$ substrate, which allows one to make a direct comparison with the report in [8] or [9]. In this paper, we present the main results. Encouragingly, the ultrathin $MgB_2$ films grown on (0001) $Al_2O_3$ substrate with thicknesses down to 7.5 nm are all found to have $T_c$ above 34 K. A high critical current density is also achieved in a series of ultrathin films from 7.5 nm thick to 40 nm thick and in particular the critical current density of 7.5 nm thick $MgB_2$ film is ~ $10^6$ A·cm$^{-2}$ at 16 K. With a much higher $T_c$ compared to films fabricated by other $MgB_2$ deposition methods such as co-evaporation, our results may restore the promise of fabricating SSPDs with ultrathin $MgB_2$ films on $Al_2O_3$ substrate.

# II. Experiment

The HPCVD technique used in this experiment has been described in detail elsewhere [12]. Mg slugs (99.5% in purity, 2 g in weight) are placed surrounding the (0001) $Al_2O_3$ substrates with square size of 5 mm × 5 mm on a molybdenum susceptor. The high Mg vapor pressure, a prerequisite for the thermodynamic stability of the $MgB_2$ phase, is realized by heating the Mg source to deposition temperatures in the range (660-720 °C). Boron is supplied by the thermal decomposition of $B_2H_6$ gas (5% in hydrogen, flow rate 1-4 sccm). Pre-purified hydrogen at 300 sccm is continuously introduced, maintaining the total system pressure at 5.2 kPa during the whole process. The thickness of the $MgB_2$ films is controlled by precisely adjusting both the $B_2H_6$ flow rate and the deposition time, and is measured by either an atomic force microscope (AFM, SPI 3800N) or a scanning electron microscope (SEM, FEI DB 235) on half-etched samples. The film topography is investigated by the above SEM. Its phase purity and crystalinity are examined by x-ray diffraction (XRD, Rigaku). The temperature dependence of resistivity is determined using a Quantum Design physical property measurement system (PPMS) with a standard four probe method employing silver paste as the contacts. The measurements of critical current density are conducted in PPMS on 5 μm × 80 μm strips which are fabricated by standard photolithography and argon milling.

## III. Results and Discussion

To accurately control the thickness of the ultrathin $MgB_2$ films, the deposition time and the $B_2H_6$ flow rate are optimized according to the SEM and AFM measurement. By keeping the other parameters constant, the $H_2$ flow rate ~ 300 sccm, total pressure ~ 5.2 kPa and the deposition temperature ~ 700 °C, a series of films with thicknesses from 7.5 to 40 nm were synthesized. The estimated thickness based on the growth conditions is consistent with the measured thickness by AFM and SEM, as presented in table 1. As can be seen, for the series of $MgB_2$ ultrathin films, the percentage deviation of the estimated thickness from the measured value is within 5%. This indicates that the deposition thickness at the above conditions basically satisfies 10 (nm·min$^{-1}$·sccm$^{-1}$) × deposition time ( min ) × $B_2H_6$ flow rate (sccm).

**Table 1.** The reaction conditions, estimated film thickness, measured thickness of $MgB_2$ films and percentage deviation of the estimated thickness from the measured thickness.

| $B_2H_6$ flow rate (sccm), deposition time (min) | Estimated value (nm) | Measured value (nm) | Percentage deviation (%) |
|---|---|---|---|
| 1, 0.50 | 5 | -- | -- |
| 1, 0.75 | 7.5 | 7.3 | 2.7 |
| 1, 1.00 | 10 | 9.7 | 3.0 |
| 2, 1.00 | 20 | 21.0 | 5.0 |
| 4, 1.00 | 40 | 41.5 | 3.8 |

The x-ray diffraction results of the series of ultrathin $MgB_2$ films are shown in figure 1. For these films, the main diffraction peaks in the scans are from the substrate and $MgB_2$ ((0001) and (0002)), indicating that there is no obvious impurity phase and the samples are predominantly $MgB_2$ with the c axis perpendicular to the substrates. These XRD results of the ultrathin films are consistent with the epitaxial growth of thick films on $Al_2O_3$ substrate reported in [13]. The basal plane of $MgB_2$ and $Al_2O_3$ possess the same hexagonal atomic configuration and six-fold symmetry. Although an a-to-a alignment between $MgB_2$ and $Al_2O_3$ results in ~ 23% lattice mismatch, which would be unfavorable for epitaxial growth, a 30° angle off a-to-a alignment provides a smaller (~ 11%) lattice mismatch that still allows an epitaxial growth [14]. Therefore, the growth relations of the epitaxial films are $[10\bar{1}0](0001)$ MgB$_2$ $\parallel$ $[11\bar{2}0](0001)$ Al$_2$O$_3$ [14] for MgB$_2$ films on Al$_2$O$_3$ substrate. The greater lattice mismatch between MgB$_2$ and Al$_2$O$_3$, compared to that between MgB$_2$ and SiC (~ 0.42%), may still result in more difficulties in fabricating MgB$_2$ ultrathin films on Al$_2$O$_3$ substrate.

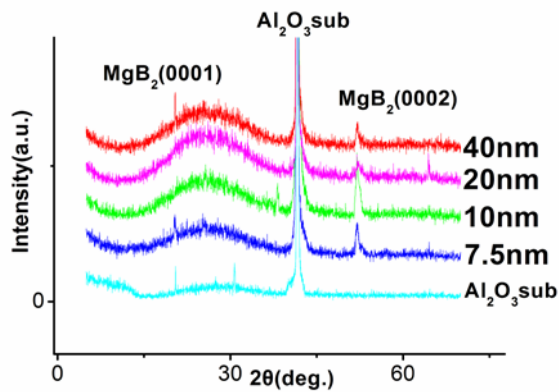

**Figure 1.** XRD scans of the Al$_2$O$_3$ substrate and MgB$_2$ films with different thickness.

The SEM images of films with thicknesses of 5, 7.5, 10 and 20 nm are shown in figure 2. The morphology of this series of films indicates the existence of island growth in the HPCVD fabrication of MgB$_2$ ultrathin films on Al$_2$O$_3$ substrate. The initial nucleation of discrete islands, as shown in the 5 nm thin film, coalesce to connected grains when the thickness increases to 7.5 nm, where the whole thin film shows zero-resistivity superconducting characteristics. For the 10 and 20 nm film, most of the grains are connected to form a continuous film, where $T_c$ is relatively unchanged. The film surface is found to be completely smooth when the film grows to 40 nm in thickness. This is consistent with our previous work reported in [13].

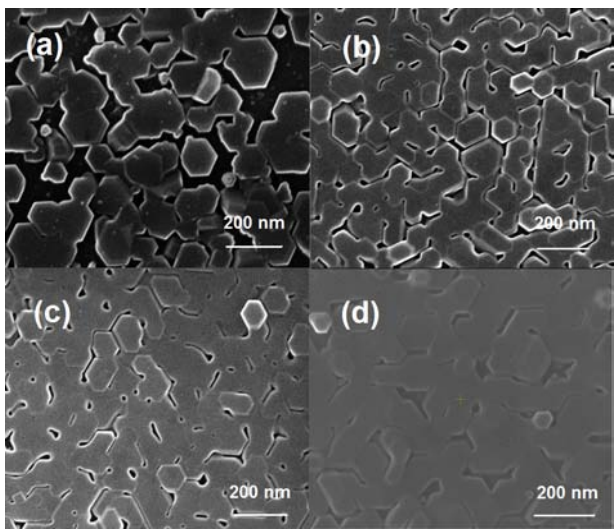

**Figure 2.** SEM images of the MgB$_2$ thin films with different thicknesses on Al$_2$O$_3$ substrates: (a) 5 nm, (b) 7.5 nm, (c) 10 nm and (d) 20 nm.

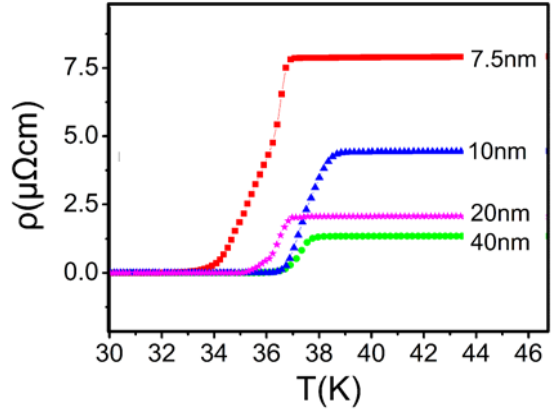

**Figure 3.** Resistivity versus temperature curves for MgB$_2$ films with thicknesses of 7.5, 10, 20 and 40 nm on Al$_2$O$_3$ substrates.

In figure 3, the resistivity versus temperature curves are presented for the series of films. For the 5 nm thin film, no superconducting transition is detected, since no connection is formed between MgB$_2$ islands, as shown in figure 2(a). The $T_c$ of 10 nm thin film and 20 nm thin film are similar due to their similar degree of the grain interconnection, as shown in figures 2(c) and (d). The $T_c$(onset), $T_c$(0) and $\rho$(42K) of the films determined from figure 3 are presented in table 2. As can be seen, the transition width, $\Delta T(K) = T_c$(onset) - $T_c$(0), decreases with the increase of the film thickness, indicating a better electronic superconducting property due to better continuity in the film surface, as shown in figure 2. Unlike the reported MgB$_2$ ultrathin films on SiC substrate, with a dramatic decrease in $T_c$(0) and a dramatic increase in $\rho$(42K) as the thickness is reduced [10], the variation of the $T_c$(0) and $\rho$(42K) of MgB$_2$ ultrathin films on Al$_2$O$_3$ substrate is smaller, as $T_c$(0) goes from 34 K for 7.5 nm thin film to 37 K for 40 nm film and $\rho$(42K) goes from 1.35 $\mu\Omega\cdot$cm for 40 nm film to 7.90 $\mu\Omega\cdot$cm for 7.5 nm thin film. At present we do not fully understand this relatively mild variation of the $T_c$(0) and $\rho$(42K), but have noted that it might relate to the existence of a very thin epitaxially grown MgB$_2$ layer below the islands shown in figure 2. This issue is under further investigation in our group. On the other hand, this lower sensitivity of the superconductivity with the decrease of film thickness is apparently advantageous for the application of MgB$_2$ ultrathin films.

**Table 2.** $T_c$(onset), $T_c$(0) and residual resistivity $\rho$(42K) of MgB$_2$ ultrathin films with different thickness derived from figure 3.

| Thickness (nm) | Transition temperature $T_c(0)$ (K) | Onset temperature $T_c(onset)$ (K) | $\Delta T(K)$ $T_c(onset) - T_c(0)$ | Residual resistivity $\rho(42K)$ ($\mu\Omega\cdot$cm) |
|---|---|---|---|---|
| 7.5 | 34.0 | 36.5 | 2.5 | 7.90 |
| 10 | 36.4 | 38.5 | 2.1 | 4.46 |
| 20 | 35.6 | 36.9 | 1.3 | 2.08 |
| 40 | 37.0 | 38.0 | 1.0 | 1.35 |

The self-field critical current density $J_c(T)$, determined by 1 $\mu$V criteria from the $I$-$V$ curves at different temperatures [11], of the series of ultrathin MgB$_2$ films are shown in figure 4. The $J_c$ (16 K) value of the 7.5 nm thin film is ~ $10^6$ A$\cdot$cm$^{-2}$, which is the first so far reported critical current density of MgB$_2$ ultrathin film down to 7.5 nm. While the $J_c(T)$ of the films on Al$_2$O$_3$ substrate are a little lower than that of SiC substrate [10, 11], the $J_c$ (20 K) of 10 nm film on Al$_2$O$_3$ substrate is nearly equal to that of 10 nm film on SiC substrate at $1 \times 10^6$ A·cm$^{-2}$. This high critical current density demonstrates the excellent crystalinity of connectivity between grains of the ultrathin MgB$_2$ films on Al$_2$O$_3$ substrate.

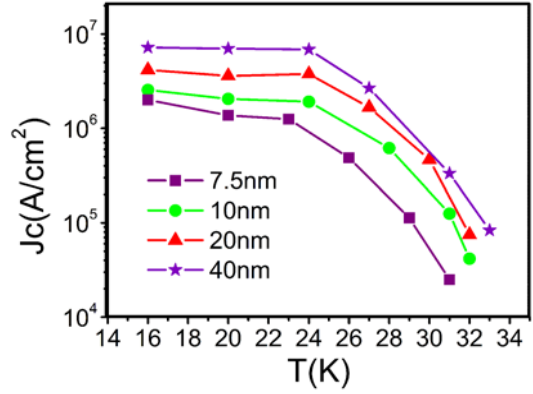

**Figure 4.** Temperature dependence of $J_c$ for MgB$_2$ films with thicknesses of 7.5, 10, 20 and 40 nm on Al$_2$O$_3$ substrates.

Until recently, there have been only a few reports on ultrathin MgB$_2$ films, including the films on Al$_2$O$_3$ substrate fabricated by co-evaporation by Shimakage *et al* [8] and molecular beam epitaxy by Shibata *et al* [9], and the films on SiC substrate fabricated by HPCVD by our group [10] and Zhuang *et al* [11]. Overcoming the difficulty of fabrication due to the relatively large mismatch between Al$_2$O$_3$ and MgB$_2$, our 7.5 nm thin film on Al$_2$O$_3$ substrate synthesized by HPCVD has $T_c$ not only much higher than the 20-22 K $T_c$ for 10 nm films by other MgB$_2$ deposition methods [8, 9], but also higher than $T_c$ of 7.5 nm film on SiC substrate [10, 11]. Our results of continuously connected 7.5-10 nm MgB$_2$ films with high $T_c$ and $J_c$ values effectively dismiss the doubt that the large decrease in $T_c$ with the decrease of film thickness may be intrinsic and unavoidable. Our results also indicate that amongst different growth techniques, HPCVD can produce rather clean and high quality ultrathin MgB$_2$ films on different kinds of substrates. This would have a positive effect on the application of MgB$_2$ in superconducting devices such as SSPDs.

## IV. Conclusion

In summary, ultrathin MgB$_2$ films down to 7.5 nm have been epitaxially grown on Al$_2$O$_3$ substrate by the HPCVD method. The 7.5 nm thin film shows a $T_c(0)$ of 34 K, a resistivity of 7.90 $\mu\Omega$·cm at 42K and a critical current density of $\sim 10^6$ A·cm$^{-2}$ at 16 K. The $T_c$ of the 7.5 nm sample is the highest so far in films with the same thickness grown on different substrates by different techniques. Our results demonstrate the excellent possibility of the application of ultrathin MgB$_2$ films on Al$_2$O$_3$ substrate in superconducting devices including SSPDs.

## Acknowledgments

This work is supported by NSFC (Nos. 50572001, 10804002, 90606023, 20731160012, and 10804003), MOST of China (973 Program, Nos. 2006CD601004, 2007CB936202/04, and 2009CB623703), NSFC/RGC (N

HKUST615/06), and the National Foundation for Fostering Talents of Basic Science under grant No. J0630311.